\documentclass[12pt,a4paper]{article}
\usepackage{graphicx}  % to include figures (can also use other packages)
\usepackage{xspace} % To avoid problems with missing or double spaces after
\usepackage{amsmath} % Adds a large collection of math symbols
\usepackage{ifthen} % for conditional statements
\usepackage{multirow}
\newboolean{pdflatex}
\setboolean{pdflatex}{true} % use this if using non-eps figures
\newboolean{articletitles}
\setboolean{articletitles}{true} % False removes titles in references

\newboolean{uprightparticles}
\setboolean{uprightparticles}{false} %Set to true to get roman particle symbols

\usepackage{amssymb}
\usepackage{amsfonts}
\usepackage{upgreek} % Adds in support for greek letters in roman typeset

\def\lhcb {\mbox{LHCb}\xspace}
\def\ux85 {\mbox{UX85}\xspace}

\ifthenelse{\boolean{uprightparticles}}%
{

 \def\Ppi         {\ensuremath{\uppi}\xspace}

 \def\PDelta      {\ensuremath{\Delta}\xspace}                 
 \def\PXi      {\ensuremath{\Xi}\xspace}                 
 \def\PLambda      {\ensuremath{\Lambda}\xspace}                 
 \def\PSigma      {\ensuremath{\Sigma}\xspace}                 
 \def\POmega      {\ensuremath{\Omega}\xspace}                 
 \def\PUpsilon      {\ensuremath{\Upsilon}\xspace}                 
 
 \def\PB      {\ensuremath{\mathrm{B}}\xspace}                 
                  
 \def\PD      {\ensuremath{\mathrm{D}}\xspace}

 \def\PK      {\ensuremath{\mathrm{K}}\xspace}

 \def\Pb      {\ensuremath{\mathrm{b}}\xspace}

 \def\Pi      {\ensuremath{\mathrm{i}}\xspace}

 \def\Ps      {\ensuremath{\mathrm{s}}\xspace}

}
{

 \def\Ppi         {\ensuremath{\pi}\xspace}

 \mathchardef\PDelta="7101
 \mathchardef\PXi="7104
 \mathchardef\PLambda="7103
 \mathchardef\PSigma="7106
 \mathchardef\POmega="710A
 \mathchardef\PUpsilon="7107
                  
 \def\PB      {\ensuremath{B}\xspace}                 
                  
 \def\PD      {\ensuremath{D}\xspace}

 \def\PK      {\ensuremath{K}\xspace}

 \def\Pb      {\ensuremath{b}\xspace}

 \def\Pi      {\ensuremath{i}\xspace}

 \def\Ps      {\ensuremath{s}\xspace}

}

\def\squark    {\ensuremath{\Ps}\xspace}

\def\bquark    {\ensuremath{\Pb}\xspace}

\def\pion  {\ensuremath{\Ppi}\xspace}

\def\kaon  {\ensuremath{\PK}\xspace}
  \def\Kbar  {\kern 0.2em\overline{\kern -0.2em \PK}{}\xspace}

\def\Kz    {\ensuremath{\kaon^0}\xspace}
\def\Kzb   {\ensuremath{\Kbar^0}\xspace}
\def\KzKzb {\ensuremath{\Kz \kern -0.16em \Kzb}\xspace}
\def\Kp    {\ensuremath{\kaon^+}\xspace}
\def\Km    {\ensuremath{\kaon^-}\xspace}

\def\KpKm  {\ensuremath{\Kp \kern -0.16em \Km}\xspace}

  \def\Dbar    {\kern 0.2em\overline{\kern -0.2em \PD}{}\xspace}
\def\D       {\ensuremath{\PD}\xspace}

\def\Dz      {\ensuremath{\D^0}\xspace}
\def\Dzb     {\ensuremath{\Dbar^0}\xspace}
\def\DzDzb   {\ensuremath{\Dz {\kern -0.16em \Dzb}}\xspace}
\def\Dp      {\ensuremath{\D^+}\xspace}
\def\Dm      {\ensuremath{\D^-}\xspace}

\def\DpDm    {\ensuremath{\Dp {\kern -0.16em \Dm}}\xspace}

\def\B       {\ensuremath{\PB}\xspace}
\def\Bbar    {\ensuremath{\kern 0.18em\overline{\kern -0.18em \PB}{}}\xspace}

\def\Bs      {\ensuremath{\B^0_\squark}\xspace}
\def\Bsb     {\ensuremath{\Bbar^0_\squark}\xspace}

  \def\Y#1S{\ensuremath{\PUpsilon{(#1S)}}\xspace}% no space before {...}!

\def\Lbar {\ensuremath{\kern 0.1em\overline{\kern -0.1em\PLambda}}\xspace}

\def\to                 {\ensuremath{\rightarrow}\xspace}

\def\CP                {\ensuremath{C\!P}\xspace}

\newcommand{\dms}{\ensuremath{\Delta m_{\squark}}\xspace}

\newcommand{\DGs}{\ensuremath{\Delta\Gamma_{\squark}}\xspace}

\newcommand{\Gs}{\ensuremath{\Gamma_{\squark}}\xspace}

\newcommand{\phis}{\ensuremath{\phi_{\squark}}\xspace}

\def\AT#1     {\ensuremath{A_{\mathrm{T}}^{#1}}\xspace}           % 2

\def\C#1      {\ensuremath{\mathcal{C}_{#1}}\xspace}                       % 9
\def\Cp#1     {\ensuremath{\mathcal{C}_{#1}^{'}}\xspace}                    % 7
\def\Ceff#1   {\ensuremath{\mathcal{C}_{#1}^{\mathrm{(eff)}}}\xspace}        % 9  
\def\Cpeff#1  {\ensuremath{\mathcal{C}_{#1}^{'\mathrm{(eff)}}}\xspace}       % 7
\def\Ope#1    {\ensuremath{\mathcal{O}_{#1}}\xspace}                       % 2
\def\Opep#1   {\ensuremath{\mathcal{O}_{#1}^{'}}\xspace}                    % 7

\newcommand{\unit}[1]{\ensuremath{\rm\,#1}\xspace}          % {kg}

\newcommand{\tev}{\ensuremath{\mathrm{\,Te\kern -0.1em V}}\xspace}
\newcommand{\gev}{\ensuremath{\mathrm{\,Ge\kern -0.1em V}}\xspace}
\newcommand{\mev}{\ensuremath{\mathrm{\,Me\kern -0.1em V}}\xspace}
\newcommand{\kev}{\ensuremath{\mathrm{\,ke\kern -0.1em V}}\xspace}
\newcommand{\ev}{\ensuremath{\mathrm{\,e\kern -0.1em V}}\xspace}
\newcommand{\gevc}{\ensuremath{{\mathrm{\,Ge\kern -0.1em V\!/}c}}\xspace}
\newcommand{\mevc}{\ensuremath{{\mathrm{\,Me\kern -0.1em V\!/}c}}\xspace}
\newcommand{\gevcc}{\ensuremath{{\mathrm{\,Ge\kern -0.1em V\!/}c^2}}\xspace}
\newcommand{\gevgevcccc}{\ensuremath{{\mathrm{\,Ge\kern -0.1em V^2\!/}c^4}}\xspace}
\newcommand{\mevcc}{\ensuremath{{\mathrm{\,Me\kern -0.1em V\!/}c^2}}\xspace}

\def\invfb   {\ensuremath{\mbox{\,fb}^{-1}}\xspace}

\def\fs   {\ensuremath{\rm \,fs}\xspace}

\newcommand{\chisq}{\ensuremath{\chi^2}\xspace}

\def\deriv {\ensuremath{\mathrm{d}\xspace}}

\def\gsim{{~\raise.15em\hbox{$>$}\kern-.85em
          \lower.35em\hbox{$\sim$}~}\xspace}
\def\lsim{{~\raise.15em\hbox{$<$}\kern-.85em
          \lower.35em\hbox{$\sim$}~}\xspace}

\def\sPlot{\mbox{\em sPlot}}

\def\pt         {\mbox{$p_{\rm T}$}\xspace}

\def\rad{\ensuremath{\rm \,rad}\xspace}

\def\evtgen     {\mbox{\textsc{EvtGen}}\xspace}
\def\pythia     {\mbox{\textsc{Pythia}}\xspace}

\def\geant      {\mbox{\textsc{Geant4}}\xspace}

\def\gauss      {\mbox{\textsc{Gauss}}\xspace}

\def\tell1  {TELL1\xspace}
\def\ukl1   {UKL1\xspace}

\newcommand{\phisphiphi}{\phis}
\newcommand{\particle}[1]{{\ensuremath{\rm #1}}}
\newcommand{\DLL}[2]{\ensuremath{\Delta  \ln{\cal L}_{\particle{#1}\particle{#2}
}}}

\newcommand{\pT}{\ensuremath{p_{\rm T}}}

\newcommand{\MeVcc}{\unit{MeV\!/\!{\it c}^2}}

\newcommand{\GeVc}{\unit{GeV\!/\!{\it c}}}

\def\fourK {\ensuremath{K \kern -0.16em K \kern -0.16em K \kern -0.16em K}}
\def\twoK {\ensuremath{K \kern -0.16em K}}

\makeatletter
\def\NAT@parse{\typeout{This is a fake Natbib command to fool Hyperref.}}
\makeatother
\usepackage[hyperfootnotes=false]{hyperref}    % Hyperlinks in references
\usepackage[all]{hypcap} % Internal hyperlinks to floats.

\usepackage{cite} % Allows for ranges in citations
\usepackage{mciteplus}

\begin{document}

%%%%%%%%%%%%%%%%%%%%%%%%%
%%%%% Title     %%%%%%%%%
%%%%%%%%%%%%%%%%%%%%%%%%%
\renewcommand{\thefootnote}{\fnsymbol{footnote}}
\setcounter{footnote}{1}

% %%%%%%% CHOOSE TITLE PAGE--------
%\onecolumn
% \input{title-LHCb-ANA}
%\input{title-LHCb-CONF}
\begin{titlepage}
\pagenumbering{roman}

% Header ---------------------------------------------------
\vspace*{-1.5cm}
\centerline{\large EUROPEAN ORGANIZATION FOR NUCLEAR RESEARCH (CERN)}
\vspace*{1.5cm}
\hspace*{-0.5cm}
\begin{tabular*}{\linewidth}{lc@{\extracolsep{\fill}}r}
\ifthenelse{\boolean{pdflatex}}% Logo format choice
{\vspace*{-2.7cm}\mbox{\!\!\!\includegraphics[width=.14\textwidth]{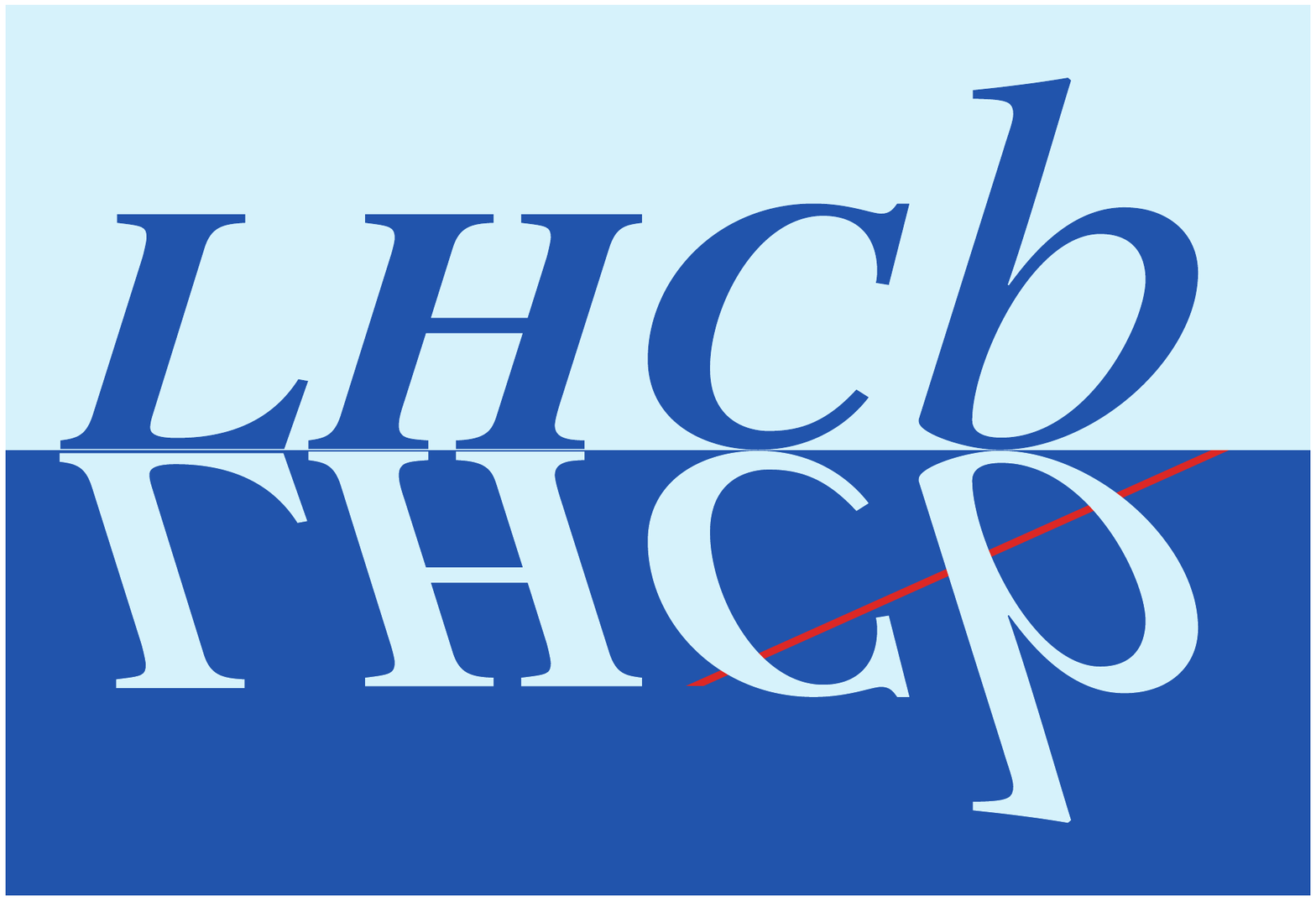}} & &}%
{\vspace*{-1.2cm}\mbox{\!\!\!\includegraphics[width=.12\textwidth]{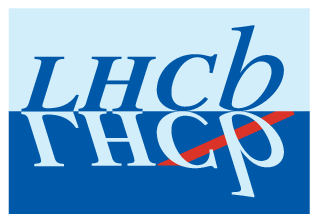}} & &}%
\\
 & & CERN-PH-EP-2013-046 \\  % ID 
 & & LHCb-PAPER-2013-007 \\  % ID 
 & & May 16, 2013 \\ % Date - Can also hardwire e.g.: 23 March 2010
 %& & v7 \\
% not in paper \hline
\end{tabular*}

%\vspace*{4.0cm}
\vspace*{2.0cm}

% Title --------------------------------------------------
{\bf\boldmath\huge
\begin{center}
  First measurement of the \CP-violating phase in $\Bs \to \phi\phi$ 
  %gluonic penguin 
  decays
\end{center}
}

\vspace*{2.0cm}

% Authors -------------------------------------------------
\begin{center}
The LHCb collaboration\footnote{Authors are listed on the following pages.}
\end{center}

\vspace{\fill}

% Abstract -----------------------------------------------
\begin{abstract}
  \noindent
  A first flavour-tagged measurement of the time-dependent \CP-violating asymmetry 
  in $\Bs \to \phi\phi$ decays is presented. In this decay channel,
  the \CP-violating weak phase arises due to \CP violation in the interference between
  \Bs-\Bsb mixing and the $b \to s \bar s s $ gluonic penguin decay
  amplitude. Using a sample of $pp$ collision data corresponding to an integrated luminosity of $1.0\;\invfb$ 
   and collected at a centre-of-mass energy of
 $7\; \rm TeV$ with the LHCb detector, $880\  \Bs \to
  \phi\phi$ signal decays are obtained. 
  The \CP-violating phase is measured to  be in the interval $\left [-2.46, -0.76 \right] \;\rm rad$ at 68\% confidence level.
  The p-value of the Standard Model prediction is $16\%$.
\end{abstract}

\vspace*{2.0cm}

\begin{center}
  Submitted to Phys. Rev. Lett.
\end{center}

\vspace{\fill}

{\footnotesize 
\centerline{\copyright~CERN on behalf of the \lhcb collaboration, license \href{http://creativecommons.org/licenses/by/3.0/}{CC-BY-3.0}.}}
\vspace*{2mm}

\end{titlepage}

%%%%%%%%%%%%%%%%%%%%%%%%%%%%%%%%
%%%%%  EOD OF TITLE PAGE  %%%%%%
%%%%%%%%%%%%%%%%%%%%%%%%%%%%%%%%

%  empty page follows the title page ----
\newpage
\setcounter{page}{2}
\mbox{~}
\newpage

% Author List ----------------------------
%  You need to get a new author list!
%%%%%%%%%%%%%%%%%%%%%%%%%%%%%%%%%%%%%%%%%%
\centerline{\large\bf LHCb collaboration}
\begin{flushleft}
\small
R.~Aaij$^{40}$, 
C.~Abellan~Beteta$^{35,n}$, 
B.~Adeva$^{36}$, 
M.~Adinolfi$^{45}$, 
C.~Adrover$^{6}$, 
A.~Affolder$^{51}$, 
Z.~Ajaltouni$^{5}$, 
J.~Albrecht$^{9}$, 
F.~Alessio$^{37}$, 
M.~Alexander$^{50}$
% ...

{\footnotesize \it
% ...
$ ^{5}$Clermont Universit\'{e}, Universit\'{e} Blaise Pascal, CNRS/IN2P3, LPC, Clermont-Ferrand, France\\
$ ^{6}$CPPM, Aix-Marseille Universit\'{e}, CNRS/IN2P3, Marseille, France\\
% ...
$ ^{9}$Fakult\"{a}t Physik, Technische Universit\"{a}t Dortmund, Dortmund, Germany\\
% ...
$ ^{35}$Universitat de Barcelona, Barcelona, Spain\\
$ ^{36}$Universidad de Santiago de Compostela, Santiago de Compostela, Spain\\
$ ^{37}$European Organization for Nuclear Research (CERN), Geneva, Switzerland\\
% ...
$ ^{40}$Nikhef National Institute for Subatomic Physics, Amsterdam, The Netherlands\\
% ...
$ ^{45}$H.H. Wills Physics Laboratory, University of Bristol, Bristol, United Kingdom\\
% ...
$ ^{50}$School of Physics and Astronomy, University of Glasgow, Glasgow, United Kingdom\\
$ ^{51}$Oliver Lodge Laboratory, University of Liverpool, Liverpool, United Kingdom\\
% ...
$ ^{n}$LIFAELS, La Salle, Universitat Ramon Llull, Barcelona, Spain\\
% ...
}
\end{flushleft}
%%%%%%%%%%%%%%%%%%%%%%%%%%%%%%%%%%%%%%%%%%

\cleardoublepage

%\twocolumn
% %%%%%%%%%%%%% ---------

\renewcommand{\thefootnote}{\arabic{footnote}}
\setcounter{footnote}{0}

%%%%%%%%%%%%%%%%%%%%%%%%%
%%%%% Main text %%%%%%%%%
%%%%%%%%%%%%%%%%%%%%%%%%%

\pagestyle{plain} % restore page numbers for the main text
\setcounter{page}{1}
\pagenumbering{arabic}

\noindent The $\Bs \to \phi\phi$  decay is forbidden at tree level in the Standard Model (SM) 
and proceeds via a gluonic $b \to s \bar s s$ penguin process. 
Hence, this channel provides an excellent probe of new heavy particles entering the penguin quantum loops~\cite{Bartsch:2008ps,beneke,PhysRevD.80.114026}.
Generally, \CP violation in the SM is governed by a single phase in the Cabibbo-Kobayashi-Maskawa quark mixing matrix
\cite{Kobayashi:1973fv,*Cabibbo:1963yz}. 
The interference between the \Bs-\Bsb oscillation and decay amplitudes leads to a \CP asymmetry
in the decay time distributions of $\Bs$ and $\Bsb$ mesons, which is characterised by a \CP-violating weak phase. 
The SM  predicts this phase to be small.
Due to different decay amplitudes  the actual value is  dependent on the $\Bs$ decay channel.
For  $\Bs \to J/\psi\phi$, which proceeds via a $b \to c \bar c s$ transition,
the SM prediction of the weak phase
is given by $- 2 \arg \left( -V_{ts} V_{tb}^*/ V_{cs} V_{cb}^*\right)=-0.036\pm0.002\; \rm rad$~\cite{Charles:2011va}.
The LHCb collaboration recently measured the weak phase in this decay 
to be $0.068 \pm 0.091 (\rm stat) \pm 0.011 (\rm syst)\; \rm rad$~\cite{LHCb-PAPER-2013-002},
which is consistent with the SM and places stringent constraints on
\CP violation in  \Bs-\Bsb
oscillations~\cite{LHCb-PAPER-2012-031}. 
In the SM, the phase in the $\Bs \to \phi\phi$ decay, \phis, is expected to be close to
zero due to a cancellation of the phases arising from  \Bs-\Bsb oscillations and decay~\cite{Raidal:2002ph}.
Calculations using QCD factorization provide an upper limit of 0.02\;\rad for $|\phisphiphi|$~\cite{Bartsch:2008ps,beneke,PhysRevD.80.114026}.

In this Letter, we present the first measurement of the \CP-violating phase in ${\Bs \to \phi\phi}$ decays.
Charge conjugate states are implied.
The result is based on  $pp$ collision data
corresponding to an integrated luminosity of  $1.0\;\invfb$ and 
 collected by the LHCb experiment in 2011 at 
a centre-of-mass energy of $7 \; \rm TeV$. 
This data sample was previously used for a time-integrated measurement of the polarisation amplitudes and triple product asymmetries 
in the same decay mode~\cite{LHCb-PAPER-2012-004}.
The analysis reported here improves the selection efficiency, measures the $\Bs$ decay time 
and identifies the flavour of the \Bs meson at production. This allows
a study of \CP violation in the interference between mixing and decay to be performed.
It is necessary to disentangle the
\CP-even longitudinal ($A_0$), \CP-even transverse ($A_\parallel$), and \CP-odd
transverse ($A_\perp$) polarisations of the $\phi\phi$ final state
by measuring the distributions of the helicity angles~\cite{LHCb-PAPER-2012-004}.

The LHCb detector is a forward spectrometer at the Large Hadron Collider covering the pseudorapidity range $2 < \eta < 5$
and is described in detail in Ref.~\cite{Alves:2008zz}.
Events are selected by a hardware trigger, 
which selects hadron or 
muon candidates with high transverse energy or momentum (\pt), followed by a two stage software trigger~\cite{Aaij:2012me}.
In the software trigger, $\Bs \to \phi\phi$ candidates are selected 
either by identifying events containing a pair of oppositely charged kaons 
with an invariant mass close to that of the  $\phi$ meson or by using a topological \bquark-hadron trigger.
In the simulation, $pp$ collisions are generated using \pythia~6.4~\cite{Sjostrand:2006za}, with a specific
LHCb configuration~\cite{LHCb-PROC-2010-056}. Decays of hadronic particles are described by \evtgen~\cite{Lange:2001uf} and
the detector response is implemented using the \geant toolkit~\cite{Allison:2006ve,*Agostinelli:2002hh} as described in Ref.~\cite{LHCb-PROC-2011-006}.

The $\Bs \to \phi \phi$ decays are reconstructed 
by combining two $\phi$ meson candidates that decay into the $K^+K^-$ final state.
Kaon candidates are required to have $\pT>0.5$\;\GeVc, and an
impact parameter (IP) $\chisq$ 
larger than 16 with respect to the primary vertex (PV),
where the IP \chisq is defined as the difference between the \chisq of the PV
reconstructed with and without the considered track.
Candidates must also be identified as kaons using the ring-imaging Cherenkov detectors~\cite{arXiv:1211-6759}, by requiring 
that the difference in the
global likelihood between the kaon and pion mass hypotheses ($\DLL{\mathit{K}}{\mathit{\pi}}\equiv\ln{\cal L}_{\mathit{K}}{}-\ln{\cal L}_{\mathit{\pi}}{}$) be larger than $-5$. 
Both $\phi$ meson candidates must have a  reconstructed mass, $m_{\twoK}$, of the kaon pair  within $20~\MeVcc$
of the known mass of the $\phi$ meson, 
a transverse momentum $(\pT^{\phi})$ larger than 0.9\;\GeVc and
a product, $\pT^{\phi 1}  \pT^{\phi 2}>2\;{\rm GeV}^2/c^2$.
The $\chisq$ per degree of freedom (ndf) of the vertex fit for both $\phi$ meson candidates and the \Bs candidate is required to be smaller than 25.
Using the above criteria, $17\,575$ candidates are selected in  the invariant four-kaon mass 
range $5100 < m_{\fourK} <  5600~\mevcc$.

A boosted decision tree (BDT)~\cite{Breiman} is used to separate signal from background.
The six observables used as input to the BDT are:
 \pt, $\eta$ and $\chisq/{\rm ndf}$ of the vertex fit for the \Bs candidate and
 the cosine of the angle between the \Bs momentum
and the direction of flight from the closest primary vertex to the decay vertex,
in addition to the smallest \pt and the largest
track  $\chisq/{\rm ndf}$ of the kaon tracks.
 The BDT is trained
using simulated $\Bs \to \phi \phi$  signal
events and background from the data
where at least one of the $\phi$ candidates has invariant mass in the range $20 < | {\it m}_{\it K \kern -0.16em K}-{\it m}_\phi | < 25\;\MeVcc$.

The \sPlot\ technique~\cite{Pivk:2004ty,2009arXiv0905.0724X}
is used to assign a signal weight to each $\Bs \to \phi \phi$  candidate.
Using the four-kaon mass as the discriminating variable,
the distributions of the signal components for the \Bs decay time and helicity angles can be determined 
in the data sample.
The sensitivity to $\phisphiphi$ is optimised taking into account the signal purity
 and the flavour tagging performance.
The final selection of  $\Bs \to \phi \phi$  candidates based on this optimisation 
is required to have a BDT output larger than 0.1,   $\DLL{\mathit{K}}{\mathit{\pi}} > -3$ for each kaon
and $|{\it m}_{\mathit K \kern  -0.16em K}-{\mathit m}_\phi | < 15\;\MeVcc$ for each $\phi$ candidate.

\begin{figure}[thpb]
\begin{center}
\includegraphics[height=8cm]{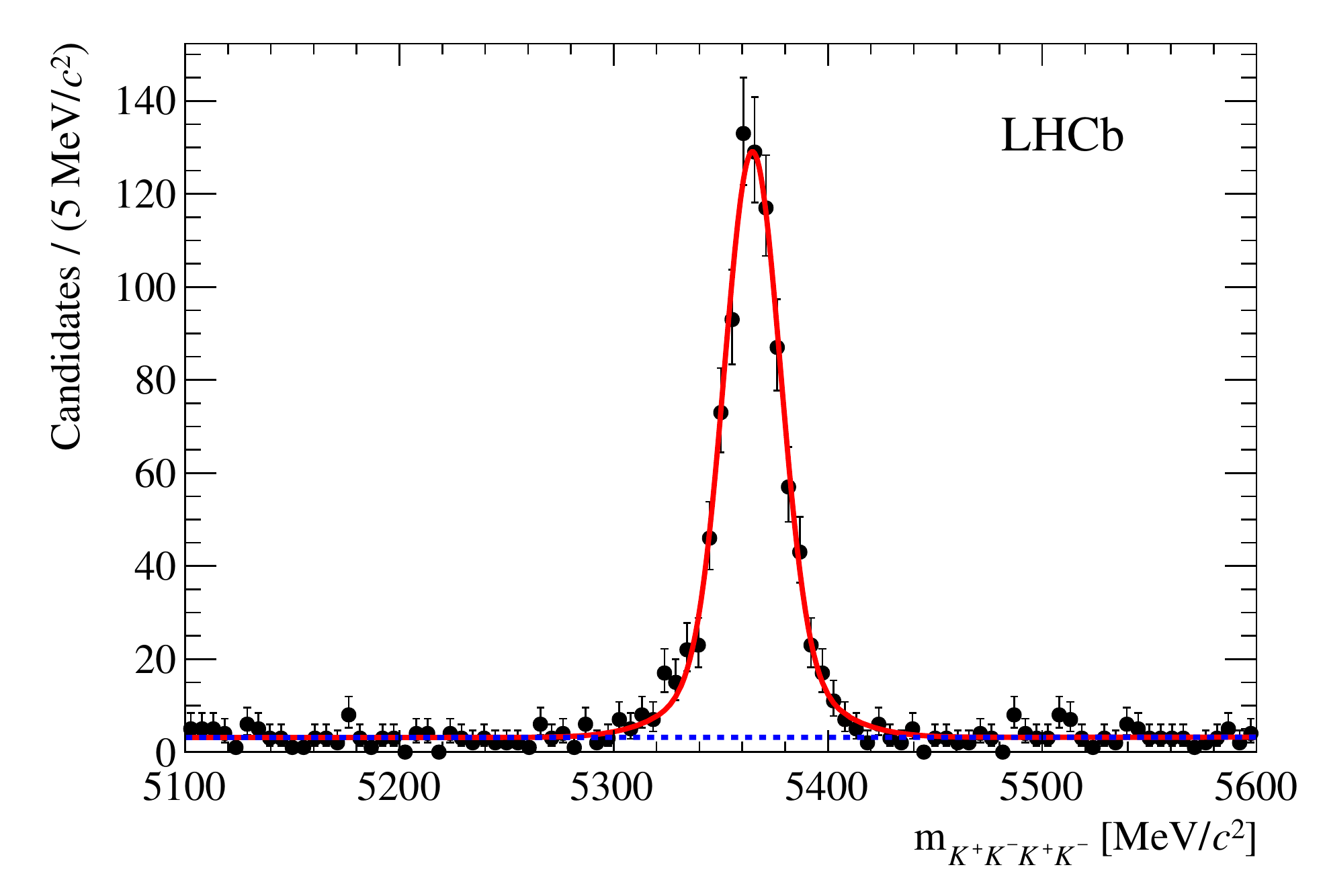}
\end{center}
\caption{\small Invariant $K^+K^-K^+K^-$ mass distribution for selected $\Bs \rightarrow \phi \phi$
  candidates. The total fit (solid line) consists
  of a double Gaussian signal component together with an exponential background
  (dotted line).}
\label{fig:mass:fourkaon}
\end{figure}

In total, 1182 $\Bs \to \phi \phi$  candidates are selected.
Figure~\ref{fig:mass:fourkaon} shows the four-kaon invariant mass distribution for the selected
events.  Using an unbinned extended maximum likelihood fit, a signal yield of
$880 \pm 31$ events is obtained. In this fit, the $\Bs \rightarrow\phi \phi$ signal component is modelled by two
Gaussian functions with a common mean. 
The width of the first Gaussian component is measured to be $12.9 \pm 0.5 \; \MeVcc$, 
in agreement with the expectation from simulation.  
The relative fraction and width
of the second Gaussian component are fixed from simulation to values of 0.785 and 29.5\mevcc, respectively,
in order to ensure a good quality fit. 
Combinatorial background is modelled 
using an exponential function which is allowed to vary in the fit. Contributions from specific backgrounds such as 
$B^0 \rightarrow \phi K^{*0}$, where $K^{*0} \to \kaon^+ \pion^-$, are found to be negligible.

An unbinned maximum likelihood fit  is performed to the decay
time, $t$, and the three helicity angles,
 $\Omega =\{\theta_1,\theta_2, \Phi\}$,
of the selected $\Bs \to \phi \phi$  candidates, each of which is re-assigned a signal  \sPlot\ weight 
based on the four-kaon invariant mass, $m_{\fourK}$\cite{Pivk:2004ty,2009arXiv0905.0724X}.
The probability density function (PDF) consists of signal components, which include detector resolution and acceptance effects,
and are factorised into separate terms for the decay time and the angular observables.

The $\Bs$ decay into the  $K^+K^-K^+K^-$ final state can proceed via combinations of  intermediate vector ($\phi$)
and scalar ($f_0(980)$) resonances  and scalar non-resonant $K^+K^-$ pairs. 
Thus the total decay amplitude is a coherent sum of $P$-wave (vector-vector),  
$S$-wave (vector-scalar) and $SS$-wave (scalar-scalar) contributions. 
The differential decay rate of the decay time and helicity angles  is described by a sum of 15 terms,
 corresponding to five polarisation amplitudes and their interference terms, 
\begin{equation}
\frac{\deriv^4 \Gamma}{\deriv \negthinspace \cos \theta_1 \, \deriv \negthinspace  \cos \theta_2 \, \deriv \Phi \, \deriv t}\propto 
\sum^{15}_{i=1}K_i(t)f_i(\Omega) \, .
\label{equ:dGdOmegadt:phiphi}
\end{equation}
The angular functions $f_i(\Omega)$ for the $P$-wave terms are derived in Ref.~\cite{PhysRevD.61.074031}
and the helicity angles of the two $\phi$ mesons are randomly assigned to $\theta_1$ and $\theta_2$.
The time-dependent functions $K_i(t)$ can be written as~\cite{PhysRevD.61.074031}
\begin{equation}
K_i(t)=N_ie^{-\Gs t} [ c_i \cos(\dms t) + d_i\sin(\dms t)
+ a_i \cosh(\tfrac{1}{2}\DGs t) + b_i \sinh(\tfrac{1}{2}\DGs t) ],
\label{equ:Ki:phiphi}
\end{equation}
where  $\DGs = \Gamma_{\rm L} - \Gamma_{\rm H}$ is the decay width difference between the 
light (L) and heavy (H) \Bs mass eigenstates, 
 $\Gs$ is the average decay width, $\Gs = (\Gamma_{\rm L }+ \Gamma_{\rm H})/2$,
 and $\dms$ is the \Bs-\Bsb oscillation frequency. 
 The coefficients $N_i$, $a_i$, $b_i$, $c_i$ and $d_i$ 
 can be expressed in terms of $\phisphiphi$ and the magnitudes, $|A_i|$, 
 and phases, $\delta_i$, of the five polarisation amplitudes at $t = 0$.
 The three $P$-wave amplitudes, denoted   by  $A_0, A_{\parallel}, A_{\perp}$,
 are normalised such that $|A_0|^2+|A_\parallel|^2+|A_\perp|^2=1$, 
with the strong phases $\delta_1$ and $\delta_2$ defined as
 $\delta_1=\delta_\perp-\delta_\parallel$ and $\delta_2=\delta_\perp-\delta_0$.
The $S$ and $SS$-wave amplitudes and their corresponding
phases are denoted by $A_S$, $A_{SS}$ and  $\delta_S$, $\delta_{SS}$, respectively.
For  a \Bs meson produced at $t = 0$,  the coefficients in Eq.~\ref{equ:Ki:phiphi} and the angular functions $f_i(\theta_1,\theta_2,\Phi)$ 
are given in Table~\ref{tab:terms},
where $\delta_{2,1} = \delta_2 - \delta_1$.
\begin{table}
\caption{\small Coefficients of the time-dependent terms and angular functions defined in Eqs.~\ref{equ:dGdOmegadt:phiphi} and \ref{equ:Ki:phiphi}. Amplitudes are defined at $t=0$.}
\begin{center}
{\scriptsize
\[
\begin{array}{c|c|c|c|c|c|c}
i     & N_i                                 & a_i         & b_i                         & c_i                         & d_i          & f_i \\ \hline
1       & |A_0|^2                          & 1             & -\cos\phis                  & 0                             & \sin\phis   & 4\cos^2\theta_1\cos^2\theta_2 \\
2       & |A_\parallel |^2                 & 1             & -\cos\phis                  & 0                             & \sin\phis   & \sin^2\theta_1\sin^2\theta_2(1{+}\cos 2\Phi) \\
3       & |A_\perp |^2                     & 1             & \cos\phis                   & 0                             & -\sin\phis  & \sin^2\theta_1\sin^2\theta_2(1{-}\cos 2\Phi) \\
4       & |A_\parallel||A_\perp |       & 0             & -\cos\delta_1\sin\phis      & \sin\delta_1      & -\cos\delta_1\cos\phis& -2\sin^2\theta_1\sin^2\theta_2\sin 2\Phi \\
5       & |A_\parallel||A_0  |          & \cos(\delta_{2,1})             & - \cos(\delta_{2,1}) \cos\phis                   & 0                             &  \cos(\delta_{2,1}) \sin\phis   & \sqrt{2}\sin2\theta_1\sin2\theta_2\cos\Phi \\
6       & |A_0 ||A_\perp|               & 0             & -\cos\delta_2\sin\phis      & \sin\delta_2      & -\cos\delta_2\cos\phis& -\sqrt{2}\sin2\theta_1\sin2\theta_2\sin\Phi \\
7       & |A_{SS}|^2                       & 1             & -\cos\phis                  & 0                             & \sin\phis   & \frac{4}{9} \\
8       & |A_S|^2                          & 1             & \cos\phis                   & 0                             & -\sin\phis  & \frac{4}{3}(\cos\theta_1+\cos\theta_2)^2 \\
9       & |A_S| |A_{SS} |               & 0             & \sin(\delta_S{-}\delta_{SS})\sin\phis & \cos(\delta_{SS}{-}\delta_{S}) & \sin(\delta_{SS}{-}\delta_{S})\cos\phis & \frac{8}{3\sqrt{3}}(\cos\theta_1+\cos\theta_2)\\
10      & |A_0 | |A_{SS}| & \cos\delta_{SS}            & - \cos\delta_{SS} \cos\phis                  & 0                             & \cos\delta_{SS}\sin\phis   & \frac{8}{3}\cos\theta_1\cos\theta_2 \\
11      & |A_\parallel | |A_{SS} | &  \cos(\delta_{2,1}{-}\delta_{SS})               & -  \cos(\delta_{2,1}{-}\delta_{SS})  \cos\phis                  & 0             &  \cos(\delta_{2,1}{-}\delta_{SS})  \sin\phis   & \frac{4\sqrt{2}}{3}\sin\theta_1\sin\theta_2\cos\Phi \\
12      & |A_\perp | |A_{SS} |          & 0             & -\cos(\delta_2-\delta_{SS})\sin\phis& \sin(\delta_2{-}\delta_{SS}) & -\cos(\delta_2{-}\delta_{SS})\cos\phis&  -\frac{4\sqrt{2}}{3}\sin\theta_1\sin\theta_2\sin\Phi \\
\multirow{2}{*}{$13$}      &\multirow{2}{*}{$ |A_0 | |A_{S} | $}               &\multirow{2}{*}{$0$}             &\multirow{2}{*}{$ -\sin\delta_S\sin\phis$}                & \multirow{2}{*}{$\cos\delta_S $}               & \multirow{2}{*}{ $ -\sin\delta_S\cos\phis $} 
& \frac{8}{\sqrt{3}} \cos\theta_1\cos\theta_2\\
& & & & & & \times (\cos\theta_1 + \cos\theta_2) \\
\multirow{2}{*}{$14$}      &\multirow{2}{*}{$ |A_\parallel | |A_{S} | $}      & \multirow{2}{*}{$0$}             & \multirow{2}{*}{$ \sin(\delta_{2,1}-\delta_S)\sin\phis $} &\multirow{2}{*}{$ \cos(\delta_{2,1}{-}\delta_S)  $}     &\multirow{2}{*}{$  \sin(\delta_{2,1}-\delta_S)\cos\phis $} 
& \frac{4\sqrt{2}}{\sqrt{3}} \sin\theta_1\sin\theta_2\\
& & & & & & \times (\cos\theta_1 + \cos\theta_2)\cos\Phi \\
\multirow{2}{*}{$15$}      & \multirow{2}{*}{$ |A_\perp | |A_{S}| $} &  \multirow{2}{*}{$ \sin(\delta_2-\delta_S) $}  & \multirow{2}{*}{$ \sin(\delta_2-\delta_S) \cos\phis $}                  & \multirow{2}{*}{0}                             & \multirow{2}{*}{$-  \sin(\delta_2-\delta_S) \sin\phis $} 
& -\frac{4\sqrt{2}}{\sqrt{3}} \sin\theta_1\sin\theta_2\\
& & & & & & \times (\cos\theta_1 + \cos\theta_2)\sin\Phi \\
\end{array}
\]
}
\end{center}
\label{tab:terms}
\end{table}
Assuming that \CP violation in mixing and direct \CP violation are negligible, the differential distribution for a $\Bsb$ meson 
is obtained by changing the sign of the coefficients $c_i$ and $d_i$. 
The PDF is invariant under the transformation 
$(\phisphiphi, \DGs, \delta_{\parallel}, \delta_{\perp}, \delta_S, \delta_{SS}) 
\rightarrow
(\pi-\phisphiphi, -\DGs, -\delta_{\parallel}, \pi-\delta_{\perp}, -\delta_S, -\delta_{SS})$.
This two-fold ambiguity is resolved in the fit as 
Gaussian constraints are applied for the \Bs average decay width and decay width difference
to the values measured in $\Bs \to J/\psi\phi$ decays,
$\Gs = 0.663 \pm 0.008\;\rm ps^{-1}$ and $\DGs = 0.100 \pm 0.017\; \rm ps^{-1}$,
with a correlation coefficient $\rho(\DGs,\Gs) =
-0.39$~\cite{LHCb-PAPER-2013-002}. 
Similarly, the \Bs oscillation frequency \dms{} is constrained to the
value $\dms = 17.73 \pm 0.05\;  \rm ps^{-1}$~\cite{LHCb-CONF-2011-050}.

A correction factor is multiplied to the interference terms in Table~\ref{tab:terms} between the  
$P$  and $S$-wave (and the $P$ and $SS$-wave) contributions to account for the 
finite $m_{KK}$ mass window considered in the amplitude integration. 
This factor is calculated from the interference between the 
different $m_{KK}$ lineshapes of the vector and scalar contributions.
The validity of the fit model has been extensively tested using simulated
data samples.

The acceptance as a function of the helicity angles is not completely
uniform due to the forward geometry of the detector and the momentum cuts
placed to the final state particles. A three-dimensional acceptance
function is determined using simulation. The acceptance factors
are included in the fit  as a normalisation of the PDF for each
of the angular terms. The acceptance function varies by less than 20\%
across the phase-space.

The event reconstruction, trigger and offline selections introduce 
a decay time dependent acceptance.
In particular for short decay times, the acceptance vanishes due to the trigger, which requires 
tracks with significant displacement from any PV.
Therefore, the decay time acceptance is determined using simulation and
incorporated by multiplying the signal PDF with a binned acceptance histogram.
The fractions of different triggers are found to be in agreement between
data and simulation.

The parameters of a double Gaussian function used to model 
the decay time resolution are determined from simulation studies.
A single Gaussian function with a resolution of $40\, \fs$ is found to have a
similar effect on physics parameters and is applied to the data fit.

The $\phisphiphi$ measurement requires that the meson flavour
be tagged as either a \Bs or $\Bsb$ meson at production.
To achieve this, both the opposite side (OS) and same side kaon (SSK) flavour
tagging methods are used~\cite{LHCb-PAPER-2011-027,LHCb-CONF-2012-033}. 
In OS tagging  the $\bar{b}$-quark hadron produced in association with the signal $b$-quark 
is exploited through the charge of a muon or electron
produced in semileptonic decays,  the charge of a kaon from a subsequent charmed hadron decay,
and the momentum-weighted charge of all tracks in an inclusively reconstructed decay vertex.
The SSK tagging makes use of kaons formed from
the  $s$-quark produced in association with the \Bs meson.
The kaon charge identifies the flavour of the signal \Bs meson.

The event-by-event mistag 
is the probability that the decision of the tagging algorithm is incorrect and
is determined by a neural network
trained on simulated events and calibrated with control samples~\cite{LHCb-PAPER-2011-027}.
The value of the event-by-event mistag is used in the fit as an observable 
and the uncertainties on the calibration parameters are propagated 
to the statistical uncertainties of the physics parameters, following the procedure 
described in Ref.~\cite{LHCb-PAPER-2013-002}. 
For events tagged by both the OS and SSK methods, a combined tagging decision is made.
The total tagging power is $\varepsilon_{tag} \mathcal{D}^2=(3.29\pm0.48)\%$, with a tagging efficiency 
of $\varepsilon_{tag}=(49.7\pm5.0)\%$ and a  dilution $\mathcal{D}=(1-2\omega)$ where $\omega$ is the average
mistag probability.
Untagged events are included in the analysis as they increase the sensitivity to $\phisphiphi$ through the $b_i$ terms
in Eq.~\ref{equ:Ki:phiphi}.

The total $S$-wave fraction is determined to be 
$(1.6\,^{+2.4}_{-1.2})\%$ 
where the  double $S$-wave contribution $A_{SS}$ is set to zero,
since the fit shows little sensitivity to $A_{SS}$. 
A fit to the two-dimensional mass, $m_{\twoK}$, for both kaon pairs,
where background is subtracted using sidebands is performed and yields
a consistent $S$-wave fraction of $(2.1\pm1.2)\%$.
%As a cross-check a sideband subtracted fit to
%the two-dimensional mass, $m_{\twoK}$, for both kaon pairs, with background subtracted using sidebands, yields a consistent  $S$-wave fraction
%of  $(2.1 \pm 1.2)\%$. 

\begin{figure}[thb]
\begin{center}
\includegraphics[width=0.49\textwidth]{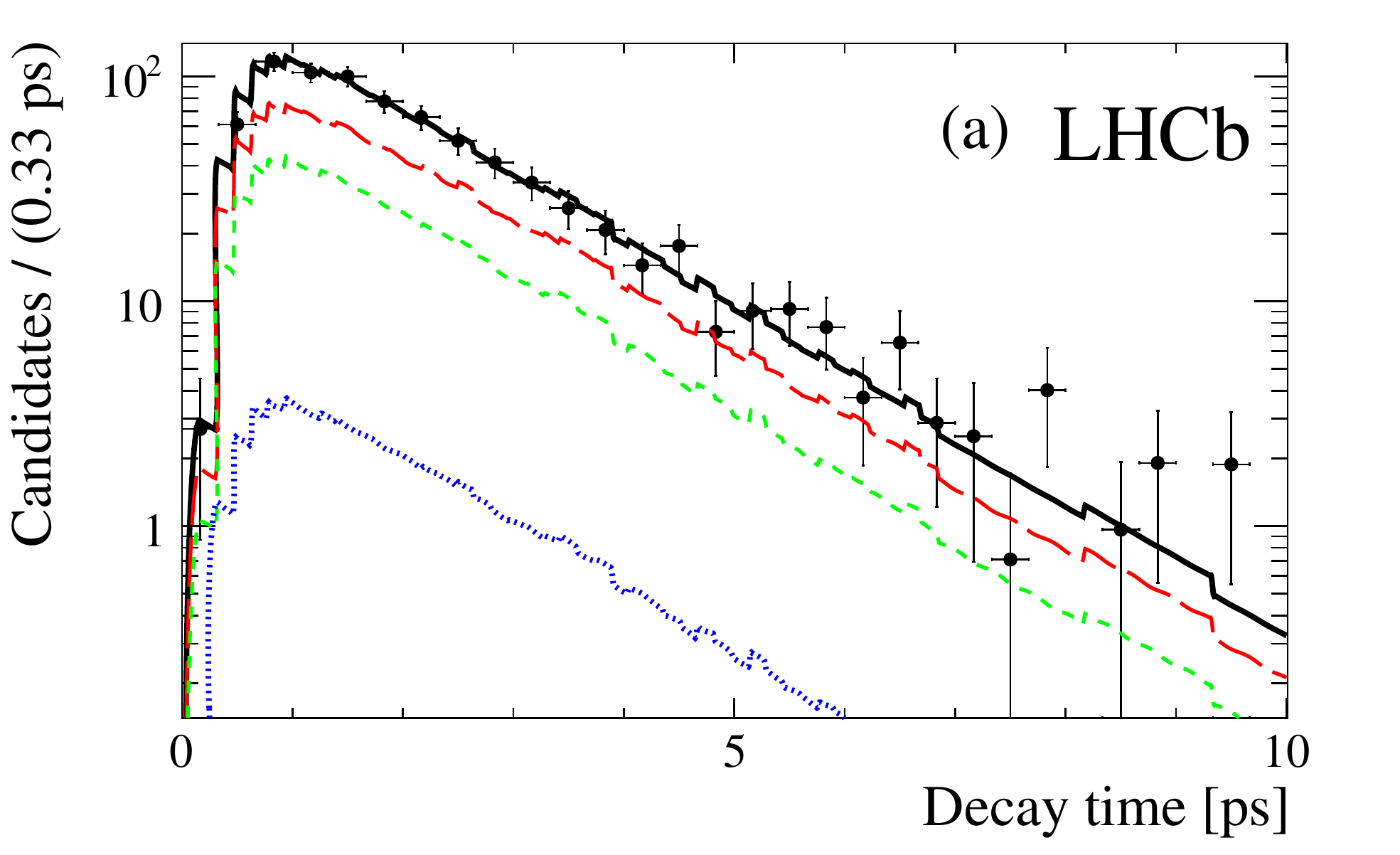}
\includegraphics[width=0.49\textwidth]{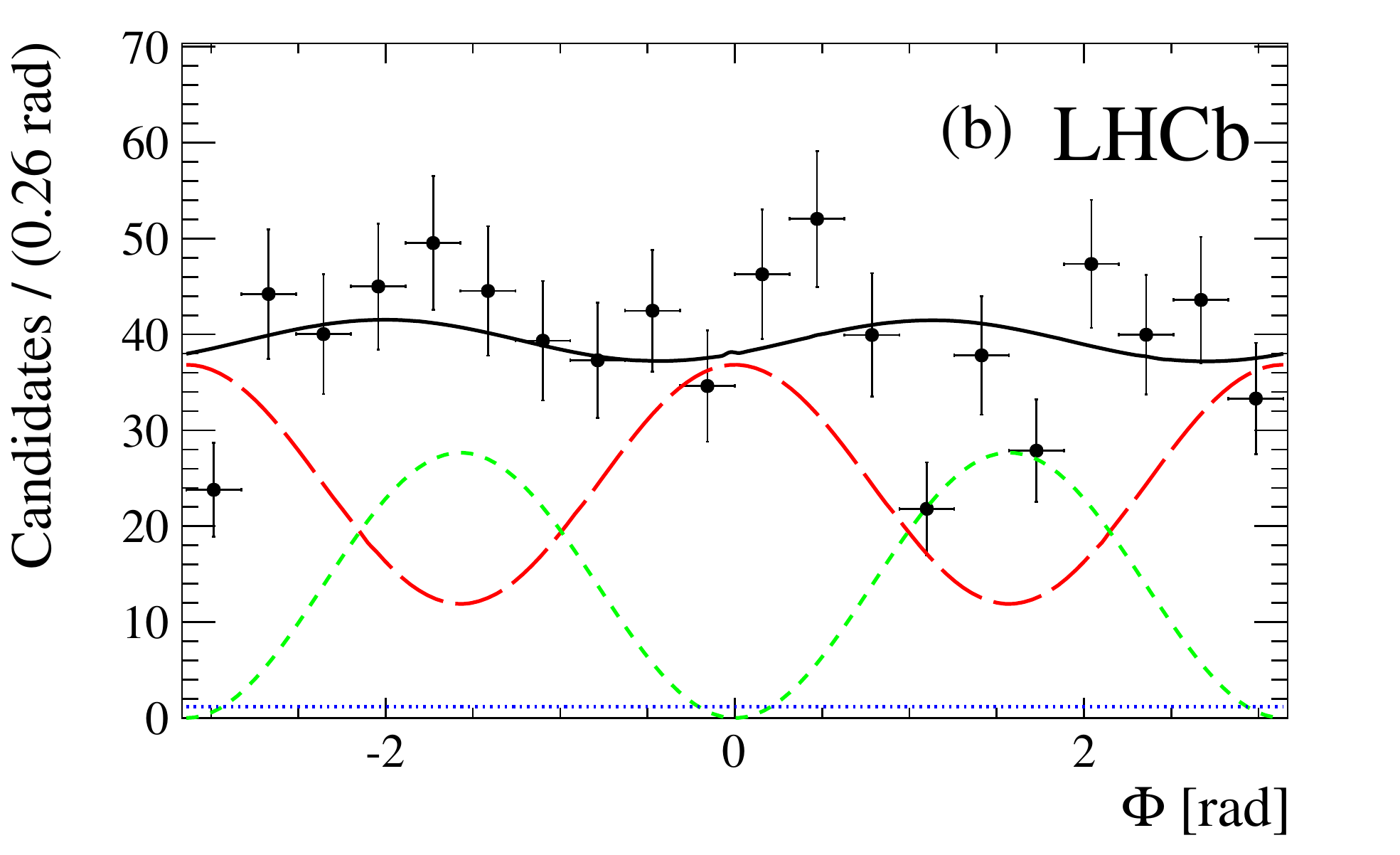} \\
\includegraphics[width=0.49\textwidth]{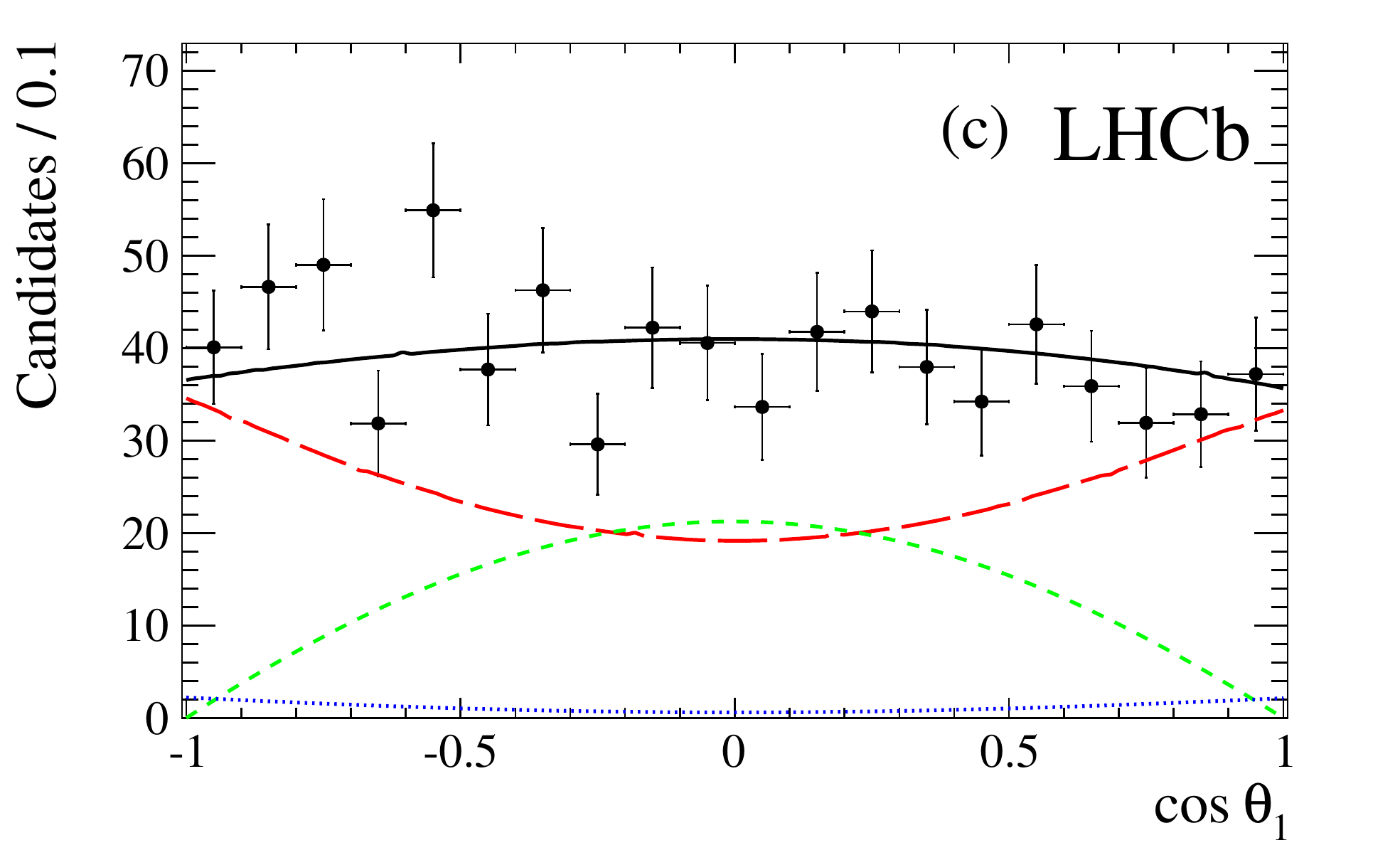}
\includegraphics[width=0.49\textwidth]{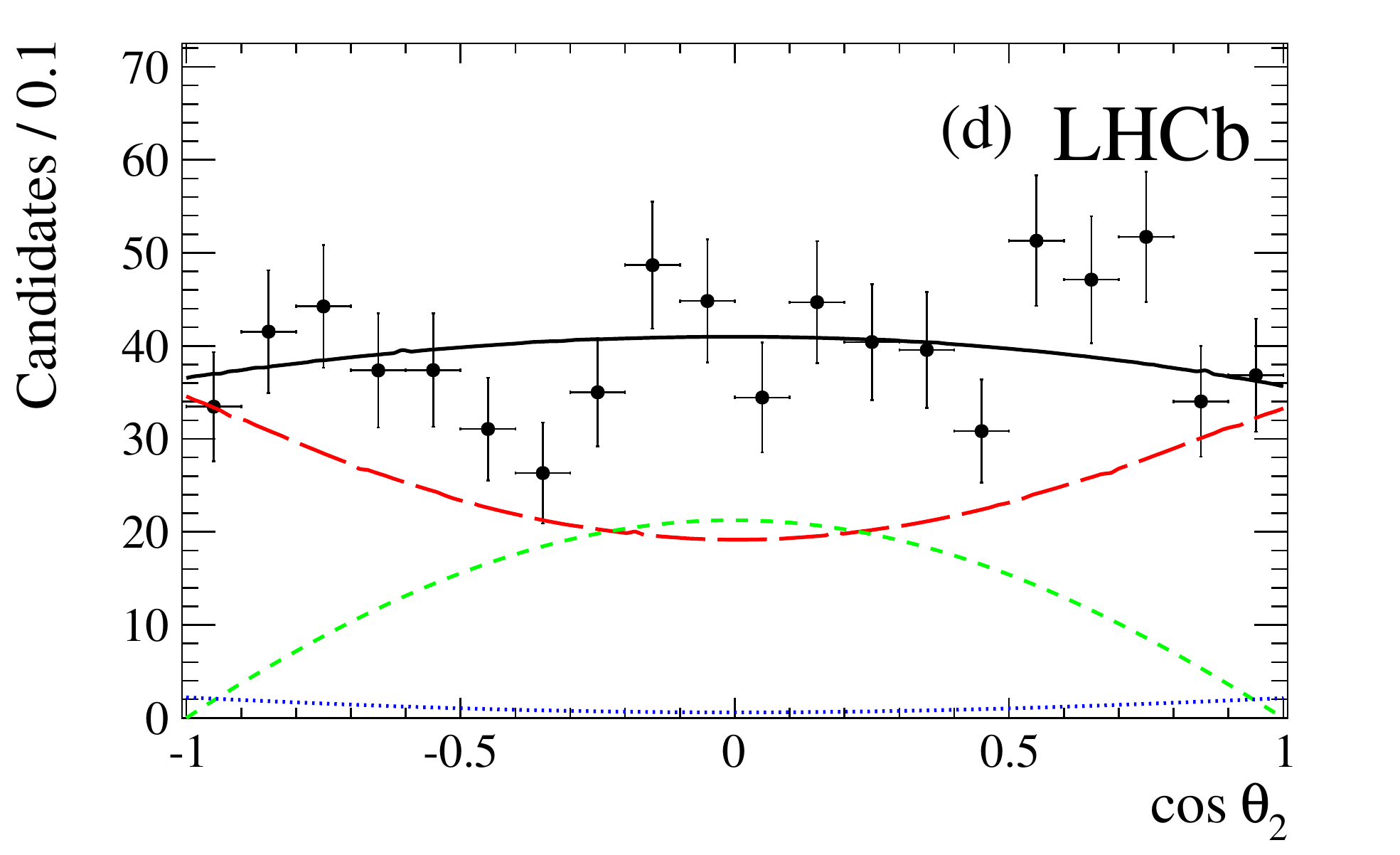}
  \caption{\small One-dimensional projections of the $\Bs \to \phi\phi$ fit for (a) decay time,  
  (b) helicity angle $\Phi$ and the cosine
  of the helicity angles (c) $\theta_1$ and (d) $\theta_2$.
   The data are marked as points, while the solid lines represent the projections of the best fit.
   The \CP-even $P$-wave, the \CP-odd $P$-wave and $S$-wave
   components are shown by the long dashed, short dashed and dotted lines, respectively.}
  \label{fig:projections}
\end{center}
\end{figure}

\begin{figure}[t]
\begin{center}
\includegraphics[height=6cm]{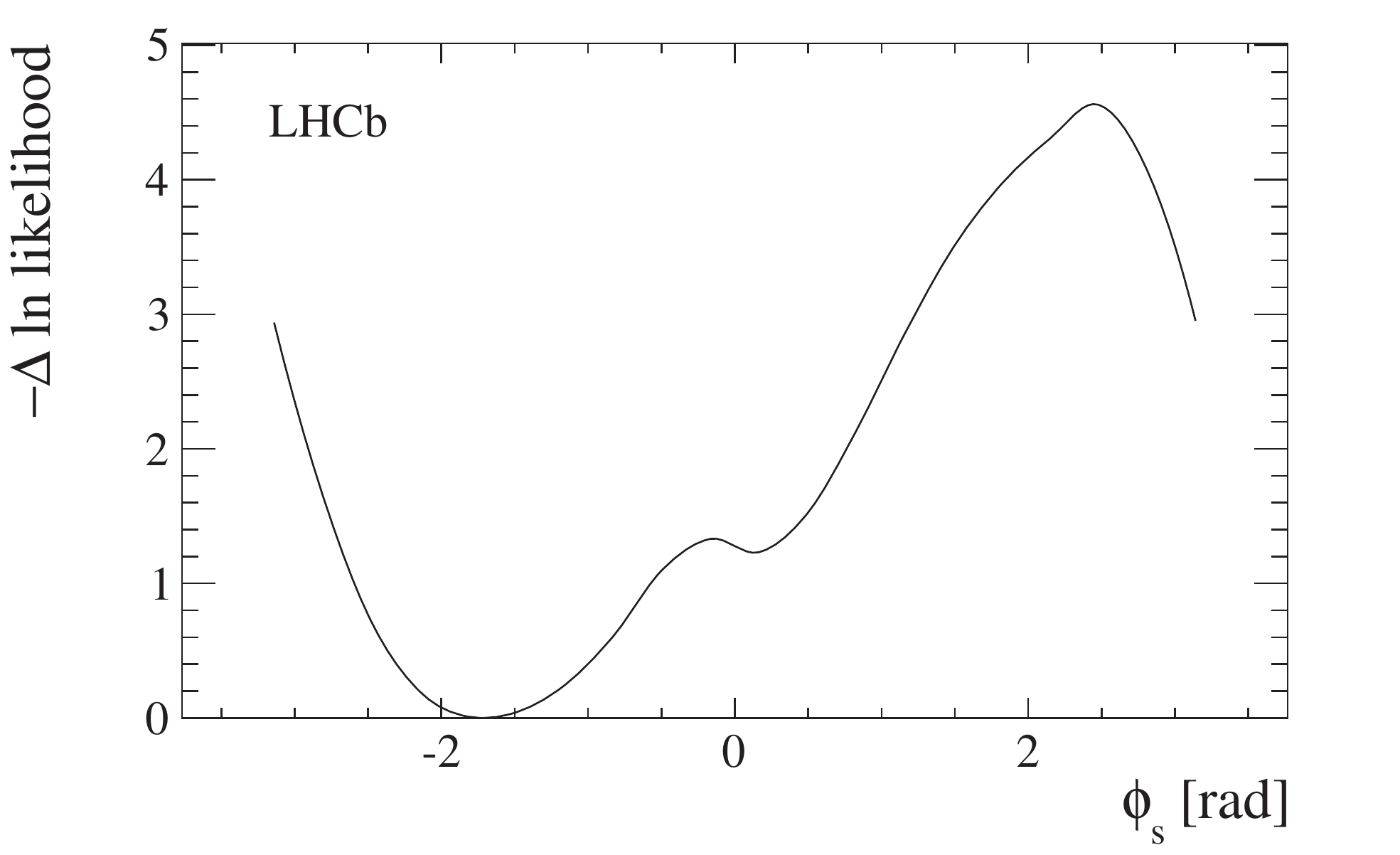}
  \caption{\small Negative $\Delta$ln likelihood scan of $\phi_s$. Only the statistical uncertainty is included.}
  \label{fig:FCLL}
\end{center}
\end{figure}

The results of the fit for the main observables are shown in Table~\ref{tab:results}. 
Figure~\ref{fig:projections} shows the distributions for the decay time and helicity angles 
with the projections for the best fit PDF overlaid.  
The likelihood profile for the \CP-violating weak phase $\phisphiphi$,  shown in Fig.~\ref{fig:FCLL}, 
is not parabolic. 
To obtain a confidence level a correction is applied due to a small under-coverage of the likelihood profile
 using the method described in Ref.~\cite{Feldman:1997qc}. 
Including systematic uncertainties (discussed below) and assuming the values of 
the polarisation amplitudes and strong phases
observed in data, an interval  of $\left [-2.46, -0.76 \right] \;\rm rad$ at  $68 \%$  confidence level
is obtained for $\phi_s$. 
The polarisation amplitudes and phases,  shown in Table~\ref{tab:results},
differ from those reported in  Ref.~\cite{LHCb-PAPER-2012-004} as $\phi_s$
is not constrained to zero.

\begin{table}[b]
{\small
  \caption{\small Fit results with statistical and systematic uncertainties. A 68\% statistical confidence interval 
  is quoted for \phis. Amplitudes are defined at $t = 0$.}
  \centerline{
    \begin{tabular*}{0.75\columnwidth}{@{\extracolsep{\fill}}@{\hspace{3pt}}p{0.2\columnwidth}@{\hspace{5pt}}ccc}
      Parameter                       & Value & $\sigma_\text{stat.}$ & $\sigma_\text{syst.}$\\ 
      \hline
     $\phi_s$[rad]  (68 \% CL)   & &           $[-2.37,-0.92]$  	& ~0.22\\ 
     $|A_{0}|^2$                  				& ~0.329 	& 0.033 	& 0.017\\
     $|A_{\perp}|^2$              				& ~0.358	& 0.046 	& 0.018\\
     $|A_\mathrm{S}|^2$      				& ~0.016 	& $^{+0.024}_{-0.012}$ 	& 0.009\\
     $\delta_{1}$     [rad]            				& ~~2.19     & ~0.44  	& ~0.12\\
     $\delta_{2}$  [rad]         					& $-1.47$  	& ~0.48 	& ~0.10\\
     $\delta_{\mathrm{S}}$     [rad]            		        &~~0.65    &   $~^{+0.89}_{-1.65}$	& ~0.33\\
    \end{tabular*}
  }
  }
  \label{tab:results}
\end{table}

The uncertainties related to the calibration of the tagging and the assumed values of
$\Gamma_s$, $\Delta \Gamma_s$ and $\Delta m_s$ are 
absorbed into the statistical uncertainty, described above.
Systematic uncertainties are determined and the sum in quadrature
of all sources is reported in Table~\ref{tab:results} for each observable.
To check that the background
is properly accounted for, an additional fit is performed where the angular and time
distributions are parameterised using the \Bs mass sidebands. This
gives results in agreement with those presented here and no
further systematic uncertainty is assigned. 
The uncertainty due to the modelling of
the $S$-wave component is evaluated by allowing the $SS$-wave 
component to vary in the fit. 
The difference between the two fits leads 
to the dominant uncertainty on $\phi_s$ of 0.20\;\rad. 
 The systematic uncertainty due to the decay time acceptance is found by 
taking the difference in the values of fitted parameters between the 
nominal fit, using a binned time acceptance, and a fit in which the time acceptance 
is explicitely paramaterised. 
This is found to be 0.09\;\rad for $\phi_s$.
  Possible differences in the
simulated  decay time resolution  compared to the data are studied by
 varying the resolution according to the discrepancies observed in
 the $\Bs \to J/\psi\phi$ analysis~\cite{LHCb-PAPER-2013-002}.
 This leads to a systematic uncertainty of 0.01\;\rad for $\phi_s$. 
 The distributions of maximum \pt and $\chisq/{\rm ndf}$ of the final state tracks and 
 the \pt and $\eta$ of the \Bs candidate are reweighted to better match the data. From this,
the angular acceptance is recalculated, leading to small changes in the results (0.02\;\rad for $\phi_s$), which are assigned
 as systematic uncertainty. 
Biases in the fit method are studied using simulated pseudo-experiments
that lead to an uncertainty  of  0.02\;\rad for $\phi_s$.
 Further small systematic uncertainties (0.02\;\rad for $\phi_s$)
 are due to the limited number of events in the simulation sample
 used for the determination of the angular acceptance and to the choice of a single versus a double Gaussian
 function for  the mass PDF, which is used to assign the signal weights.
 The total systematic uncertainty on $\phi_s$ is 0.22\;\rad, significantly
 smaller than the statistical uncertainty.
 
In summary, we present the first study of \CP violation in the decay
time distribution of hadronic $\Bs \to \phi\phi$ decays. The \CP-violating phase, $\phi_s$, is restricted to the
 interval of $\left [-2.46, -0.76 \right] \;\rm rad$ at 68\% C.L. 
The p-value of the Standard Model prediction~\cite{Raidal:2002ph} is 16\%, taking the values of the
strong phases and polarisation amplitudes observed in data and assuming that systematic uncertainties are
negligible.
The precision of the $\phi_s$ measurement is dominated by the statistical uncertainty and is expected to
improve with larger LHCb data sets.

\section*{Acknowledgements}

\noindent We express our gratitude to our colleagues in the CERN
accelerator departments for the excellent performance of the LHC. We
thank the technical and administrative staff at the LHCb
institutes. We acknowledge support from CERN and from the national
agencies: CAPES, CNPq, FAPERJ and FINEP (Brazil); NSFC (China);
CNRS/IN2P3 and Region Auvergne (France); BMBF, DFG, HGF and MPG
(Germany); SFI (Ireland); INFN (Italy); FOM and NWO (The Netherlands);
SCSR (Poland); ANCS/IFA (Romania); MinES, Rosatom, RFBR and NRC
``Kurchatov Institute'' (Russia); MinECo, XuntaGal and GENCAT (Spain);
SNSF and SER (Switzerland); NAS Ukraine (Ukraine); STFC (United
Kingdom); NSF (USA). We also acknowledge the support received from the
ERC under FP7. The Tier1 computing centres are supported by IN2P3
(France), KIT and BMBF (Germany), INFN (Italy), NWO and SURF (The
Netherlands), PIC (Spain), GridPP (United Kingdom). We are thankful
for the computing resources put at our disposal by Yandex LLC
(Russia), as well as to the communities behind the multiple open
source software packages that we depend on.

\ifx\mcitethebibliography\mciteundefinedmacro
\PackageError{LHCb.bst}{mciteplus.sty has not been loaded}
{This bibstyle requires the use of the mciteplus package.}\fi
\providecommand{\href}[2]{#2}

\end{document}